\documentclass[twocolumn,aps,prl,superscriptaddress]{revtex4}
\usepackage{graphicx}
\usepackage{amsfonts}
\usepackage[varg]{txfonts}
\usepackage{bm}

\begin{document}

\title{Fermionization Transform for Certain Higher-Dimensional Quantum Spin Models}
\author{Victor Galitski}
\affiliation{Department of Physics and Joint Quantum Institute,
University of Maryland, College Park, MD 20742-4111}

\begin{abstract}
Proposed is a generalization of Jordan-Wigner transform that allows to exactly fermionize
a large family of quantum spin Hamiltonians in dimensions higher than one.
The key new steps are to enlarge the Hilbert space of the original model by adding to it a collection of stand-alone free spins and to  use
a combination of these auxiliary operators and the lattice spins to construct a proper fermion representation of the physical Hamiltonian.
The transform is especially useful for lattice spin Hamiltonians, where two-spin interactions of $XY$-type are either absent or 
exist only within one-dimensional chains and where the chains are coupled via two-spin interactions of Ising type, ring-exchange terms, or more general 
multi-spin interactions that involve an even number of spin operators from each chain.  Using the proposed fermionization method we
provide a simple argument suggesting that a spin Hamiltonian closely-related to the ring-exchange model  proposed
by Paramekanti, Balents, and Fisher [Phys. Rev. B {\bf 66}, 054526 (2002)] indeed realizes a spin-liquid state.
\end{abstract}

%\pacs{71.10.-w, 71.10.Ay, 74.72.-h}
\maketitle

Quantum magnetism is one of the richest areas of research in condensed matter physics. It is also one of the most complicated and least understood subjects, primarily because most quantum magnets represent  strongly correlated systems, which do not necessarily relate to a simple non-interacting model that could be studied using perturbative techniques. The phases and low-energy excitations of quantum magnets are non-universal and much is determined
by  energetics. Some insight can be obtained using effective field theories~\cite{AA_book}, which  represent  low-energy physics of various low-temperature magnetic phases, but to obtain exact unambiguous results is rarely possible.

One-dimensional quantum magnets stand out as a notable exception to this state of affairs. In many cases, quantum spin chains  can be solved exactly using various non-perturbative methods~\cite{Giamarchi_book}. Perhaps, the simplest such example is a chain of half-integer spins with nearest-neighbor $XY$ interactions, {\em i.e.}, $\hat{\cal H}_{\rm XY,\,1d} = -J \sum_n \left[ \hat{\sigma}_n^+ \hat{\sigma}_{n+1}^- +{\rm h.c.}\right]$. The nature of the ground state of this spin model can be understood with the help of Jordan-Wigner transformation, proposed back in 1928~\cite{JW}. Since our Letter intends to generalize it to some higher-dimensional models, let us reiterate the key general ideas, which of course are well-known: To solve  a quantum spin Hamiltonian implies to either calculate the partition function or to prove that it is equivalent to a partition function associated with a different Hamiltonian, which we understand well, preferably a Hamiltonian expressed in terms of creation/annihilation operators of some canonical fermions or bosons. The main difficulty in accomplishing this task is due to the fact that the spin operators, $\hat{\sigma}^{\pm}_n = \left( \hat{\sigma}^x_n \pm i \hat{\sigma}^y_n \right)/2$, are neither fermions nor bosons, because they commute on different sites $\left[ \hat{\sigma}^{a}_n, \hat{\sigma}^{b}_m \right]_- \propto \delta_{nm}$ and anti-commute on the same site: $\left[ \hat{\sigma}^{a}_n, \hat{\sigma}^{b}_n \right]_+ =0$  (here and below Latin indices, $a,b = \pm$). Clearly, no local transformation can ``correct'' the anticommutation relation, but Jordan and Wigner showed that there exists a non-local transform, now bearing their names,
that accomplishes just that~\cite{JW}
\begin{equation}
\label{JW}
\hat{f}^\dagger_n = \hat{\sigma}^{+}_n \prod_{m < n} \hat{\sigma}^{z}_m \mbox{  and  }
\hat{f}_n = \hat{\sigma}^{-}_n \prod_{m < n} \hat{\sigma}^{z}_m,
\end{equation}
where $\hat{f}_n^\dagger$ and $\hat{f}_n$ are creation/annihilation fermion operators,  $\left[ \hat{f}_n, \hat{f}^{\dagger}_m \right]_+ =\delta_{nm}$. Since $\left( \hat{\sigma}^z_n \right)^2 =\hat{1}$,  $\left( \hat{\sigma}^a_n \right)^2 =0$, and $\hat{\sigma}^z_n = 2 \hat{\sigma}^+_n\hat{\sigma}^-_n - \hat{1}$, the Hamiltonian for the $XY$-spin chain with nearest-neighbor interactions becomes that of a one-dimensional Fermi gas in the language of $f$-fermions (\ref{JW}). The Heisenberg spin chain, or a more general $XYZ$-spin chain with nearest-neighbor interactions, take the form of  interacting fermion models, which we recognize as a Luttinger liquid.   These mappings are possible  because the infinite products or strings in Eq.~(\ref{JW}) collapse into unity operators in all  terms of these Hamiltonians.

 It is clear that  transform (\ref{JW}) itself  is independent of the underlying spin model and it always produces proper fermion operators out of the original spins. The problem however arises when we use these fermions to describe a generic spin Hamiltonian, {\em e.g.}, a higher-dimensional lattice model~\cite{Fradkin_CS} or a spin chain with longer-range interactions: The string products there generally do not disappear and lead to a Hamiltonian, which features terms $\propto \hat{f}_n^\dagger\exp\left[ i\pi \sum\limits_{m \in C_n} \hat{f}_m^\dagger\hat{f}_m\right]$, where $C_n$ is a parameterization- and model-dependent set of sites surrounding a site $n$. These remaining strings lead in general to various fermion-gauge-theories with constraints. These are complicated models that we generally do not know how to solve and this renders the usefulness of the canonic Jordan-Wigner fermionization doubtful. In this Letter, we suggest that by enlarging the Hilbert space of the original spin model to include ``external'' auxiliary states, one can fermionize a large class of quantum spin models in such a way that both the gauge/string factors and the new auxiliary operators disappear from the resulting fermion Hamiltonian, which therefore takes the familiar form of an interacting fermion model. Before proceeding to specific examples, let us first present the general idea of this construction.

 Consider a lattice spin Hamiltonian, $\hat{\cal H}_{s} \left[ \{ \hat{\sigma}_n^a \} \right]$,  expressed in terms of the Pauli matrices, which satisfy the familiar commutation relations, $\left[ \hat{\sigma}^\alpha_n, \hat{\sigma}^\beta_m \right] =2 i \varepsilon_{\alpha \beta \gamma} \hat{\sigma}^\gamma_n  \delta_{nm}$.  Let us also include in the model another Hamiltonian, $\hat{\cal H}_{\tau} \left[ \{ \hat{\tau}_l\} \right]$, which is expressed in terms of some other operators, $\hat{\tau}$, that commute with all lattice spins and that belong to a representation of an algebra, which may or may not be $\mathfrak{su}(2)_k$ (in all examples below, we consider  spin-one-half systems only). The combined Hamiltonian therefore is
\begin{equation}
\label{Hs+Ht}
\hat{\cal H} = {\cal H}_{s} \left[ \{ \hat{\sigma}_n^a \} \right] + {\cal H}_{\tau} \left[ \left\{ {\hat{\tau}}_l \right\} \right].
\end{equation}
The main idea here is to use {\em both} the original spins {\em and} the operators of the auxiliary algebra to construct fermion operators $\hat{f}_n = F \left[ \{ \hat{\sigma}_n^a \} ,\left\{ {\hat{\tau}}_l \right\} \right]$ with the goal of achieving the simplest possible form of the resulting fermion theory. Furthermore, since the auxiliary algebra may be needed only to ``correct'' the commutation relations, the actual Hamiltonian, ${\cal H}_{\tau}$, in Eq.~(\ref{Hs+Ht}) can be assumed to take the simplest possible form, which quite generally implies that ${\cal H}_{\tau} = 0$. Hence the partition function of the original spin model  $Z_s ={\rm Tr}\, \exp\left( - \beta \hat{\cal H}_{s} \right)$ is simply given by $Z_s = Z/Z_\tau$, where $Z_\tau =  d_{\tau}$ is the dimensionality of the auxiliary Hilbert space, which is an irrelevant constant independent of any parameters of the physical model or temperature. On the other hand, a  fermionization transform (unspecified at this stage) would lead to the Hamiltonian,
\begin{equation}
\label{Hs+0}
\hat{\cal H} = \tilde{\cal H}_{s} \left[ \{ \hat{f}_n^\dagger \},\{ \hat{f}_n \},\left\{ {\hat{\tau}}_l \right\} \right] + 0.
\end{equation}
We  show below that there exists a large class of quantum spin models that can be conveniently fermionized  with the help of auxiliary ``external'' operators, associated with free stand-alone spins [while  the Jordan-Wigner transform (\ref{JW}) that involves only ``internal'' spins is generally not as useful].

To provide the simplest explicit example, let us consider first a two-leg spin ladder with the following Hamiltonian
\begin{eqnarray}
\label{XY2leg}
\nonumber
\hat{\cal H}_{\rm XY,2\, {\rm  leg}} =  &&\!\!\!\!  \sum\limits_{n;l=1,2} \left\{
 \left[J_{||}(l) \hat{\sigma}^+(n,l) \hat{\sigma}^-(n+1,l) +{\rm h.c.}\right] - B \hat{\sigma}^z(n,l) \right\}\\
\nonumber
 \\
&&\!\!\!\! +  \sum\limits_{n,m; l_1 \ne l_2}   \left[J_{z}(n,m) \hat{\sigma}^z(n,l_1) \hat{\sigma}^z(m,l_2) +{\rm h.c.}\right].
\end{eqnarray}
Eq.~(\ref{XY2leg}) describes two $XY$-spin chains in a magnetic field, coupled via Ising-type interactions.
Let us now introduce an auxiliary spin-$1/2$ particle, $\hat{\bm \tau}$, with zero Hamiltonian and define the following composite operators
\begin{eqnarray}
\label{2legferm}
\nonumber
&& \hat{f}^\pm_\uparrow (n) = \hat{\sigma}^{\pm}(n,1) \prod_{m < n} \left[ \hat{\sigma}^{z}(m,1) \right] \hat{\tau}_x \mbox{  and }\\
&& \hat{f}^\pm_\downarrow (n) = \hat{\sigma}^{\pm}(n,2) \prod_{m < n} \left[ \hat{\sigma}^{z}(m,2) \right] \hat{\tau}_y,
\end{eqnarray}
where here and below we identify $\hat{f}^- \equiv \hat{f}$ and $\hat{f}^+ \equiv \hat{f}^\dagger = \left( \hat{f}^- \right)^\dagger$.
Since $\left[\hat{\sigma}^{z}(n,l)\right]^2 = \left[\hat\tau^\alpha\right]^2 = 1$, the inverse transform is simply
$\hat{\sigma}^{\pm}(n,1\,\,/\,\,2) =  \hat{f}^\pm_{\uparrow/\downarrow} (n) \prod_{m < n} \left[ \hat{\sigma}^{z}(m,1\,\,/\,\,2) \right] \hat{\tau}_{x\,/\,y}$.
The infinite products in Eq.~(\ref{2legferm}) represent the familiar Jordan-Wigner strings (\ref{JW}) that give rise to the desired anticommutation relations within the chains, while the last factor restores proper fermion algebra for all operators involved, {\em i.e.}
$$
\left[ \hat{f}^{a_1}_{l_1} (n_1), \hat{f}^{a_2}_{l_2} (n_2) \right]_+ = \delta_{n_1 n_2} \delta_{l_1 l_2} \delta_{a_1, -a_2},
$$
where $n_{1,2} \in \mathbb{Z}$ labels sites,  $l_{1,2} = 1,2 \equiv \uparrow,\downarrow$ labels legs,
and $a_{1,2} = \pm$ distinguishes creation and annihilation operators.
Furthermore, since $\hat{\sigma}^z (n,l) = 2 \hat{\sigma}^+(n,l) \hat{\sigma}^-(n,l) - 1 = 2  \hat{f}^{\dagger}_{l} (n) \hat{f}_{l} (n) -1, \forall n, l$, the fermion representation of (\ref{XY2leg}) does not involve the auxiliary $\tau$-operators ({\em c.f.}, Refs.~[\onlinecite{JW_ladders1},~\onlinecite{JW_ladders2}]):
\begin{eqnarray}
\label{F2leg}
\nonumber
 &&\!\!\!\! \hat{\tilde{\cal H}}_{\rm XY,2\, {\rm  leg}} =   -\sum\limits_{n;l=\uparrow,\downarrow} \left\{
 \left[J_{||}(l) \hat{f}^\dagger_l(n) \hat{f}_l(n+1) +{\rm h.c.}\right] + B \hat{f}^\dagger_l(n) \hat{f}_l(n)  \right\}\\
&&\!\!\!\! +  \sum\limits_{n,m}   \left\{J_{z}(n,m) \left[ 2 \hat{f}^\dagger_\uparrow (n) \hat{f}_\uparrow (n) - 1 \right]
\left[ 2 \hat{f}^\dagger_\downarrow (m) \hat{f}_\downarrow (m) - 1 \right]  +{\rm h.c.}\right\},
\end{eqnarray}
where we have omitted an unimportant constant. We see  that transform (\ref{2legferm}) maps the spin model onto that of two species of fermions with interactions. Let us note here in passing that the inclusion of ferromagnetic couplings would translate in Eq.~(\ref{F2leg}) into an attractive interaction between the fermions, which is expected to induce pairing correlations at $T \to 0$. In particular, the nearest-neighbor intra-chain attraction would select an $s$-wave pairing, while a combination of nearest-neighbor and zigzag  intra-chain  couplings (\ref{XY2leg}) are expected to give rise to a $p$-wave Cooper pairing~\cite{CSVGSDS} for fermions in Eq.~(\ref{F2leg}). The zero-temperature ground state of the latter model may be related to a one-dimensional topological superconductor. The spinless version of this phase was in fact one of the first models proposed by Kitaev~\cite{Kitaev} as a realization of a topological qubit, which would be associated here with the  Jackiw-Rebbi/Majorana modes~\cite{JR} at the ladder boundaries.

The proposed fermionization scheme can be generalized to more complicated Hamiltonians and in particular to higher-dimensional lattice spin models. The only requirement is that in order for the auxiliary operators to disappear from the resulting fermion Hamiltonian, the original spin interactions should involve an even number of spin operators from each leg. Note also that a ``leg''  does not necessarily have to be an actual row or column of a lattice, these legs may be formed by any one-dimensional paths, as long as the collection of these paths covers the entire lattice and no two paths intersect.

Let us now take the next simplest step and consider a periodically-translated two-leg spin ladder (\ref{XY2leg}) to form a two-dimensional square lattice consisting of coupled $XY$-spin chains. We now define the following fermion operators associated with the lattice sites, ${\bf r} = (n,l)$:
\begin{eqnarray}
\label{theferm}
\nonumber
&& \hat{f}^\dagger (n,l) = \hat{\sigma}^{+}(n,l) \prod_{m < n} \left[ \hat{\sigma}^{z}(m,l) \right]
\prod_{k < l} \left( \hat{\tau}^{y}_{k} \right) \hat{\tau}^{x}_l \mbox{  and }\\
&& \hat{f} (n,l) = \hat{\sigma}^{-}(n,l) \prod_{m < n} \left[ \hat{\sigma}^{z}(m,l) \right]
\prod_{k < l} \left( \hat{\tau}^{y}_k \right) \hat{\tau}^{x}_l,
\end{eqnarray}
where, we introduced a spin-$1/2$ operator for each leg [labeled in Eq.~(\ref{theferm}) by indices $k$ and $l$]. The existence of these new operators
does not disturb the anticommutation relations within each chain, because for any product $\hat{f}^{a_1} (n_1,l) \hat{f}^{a_2} (n_2,l)$ the ``new'' $\tau$-strings square up to an identity operator, while the ``old'' $\sigma$-strings enforce the usual constraints. However, the existence of the $\tau$-operators is crucial to produce the right anticommutation relations for operators in different legs, {\em e.g.}, for $l_1 < l_2$, we obtain:
\begin{eqnarray}
\label{acomm_check}
&& \!\!\!\!\!\!\!\!\!\!\!\! \left[ \hat{f}^{a_1}(n_1,l_1), \hat{f}^{a_2} (n_2,l_2) \right]_+ = \hat{\sigma}^{a_1}(n_1,l_1) \prod_{m_1 < n_1} \left[ \hat{\sigma}^{z}(m_1,l_1) \right] \\
&\times& \!\!\! \hat{\sigma}^{a_2}(n_2,l_2) \prod_{m_2 < n_2} \left[ \hat{\sigma}^{z}(m_2,l_2) \right]
\prod_{l_1 < k < l_2} \left( \hat{\tau}^{y}_k \right) \hat{\tau}^{x}_{l_2} \left[  \hat{\tau}^{x}_{l_1}, \hat{\tau}^{y}_{l_1} \right]_+ = 0.
\nonumber
\end{eqnarray}
Therefore, operators (\ref{theferm}) are indeed fermions. Per the same arguments as in the two-leg-spin-ladder case (\ref{XY2leg}), we find that its two-dimensional realization can be fermionized via (\ref{theferm}) in such a way that no Jordan-Wigner strings, nor $\tau$-operators appear in the Hamiltonian: We simply can replace $ \hat{\sigma}^{+}(n,l) \to  \hat{f}^\dagger (n,l)$, $\hat{\sigma}^{-}(n,l) \to  \hat{f} (n,l)$, $\hat{\sigma}^{z}(n,l) = \left[2\hat{f}^\dagger (n,l) \hat{f} (n,l) - 1 \right]$, and change the sign of the $XY$-interaction term to account for $\hat{\sigma}^{+}(n,l)\hat{\sigma}^{z}(n,l) = -\hat{\sigma}^{+}(n,l)$.

%%%%%%%%%%%%%%%%%%%%%%%%%%%%
\begin{figure}[t]
\begin{center}
\includegraphics[width=0.45\textwidth]{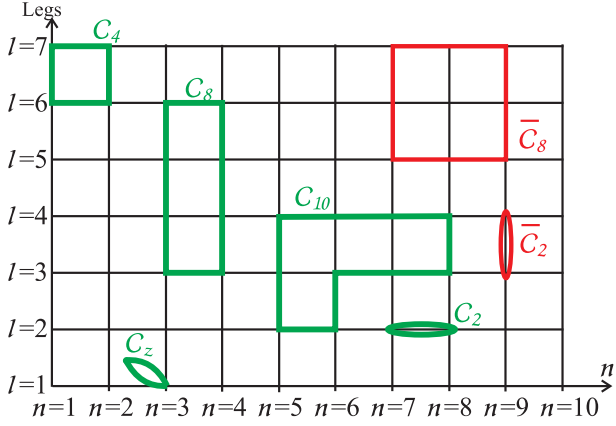}
\end{center}
\caption{(Color online):~Displayed is a square lattice associated with the general spin model (\protect\ref{GenH}) and also shown are allowed and forbidden cycles on the lattice (depicted as green and red loops correspondingly). Each loop represents a spin interaction term in the Hamiltonian, which involves a product of $L$ $\hat{\sigma}^{\pm}({\bf r})$-operators, where $L$ is the number of sites in the loop, ${\cal C}_L$. Shown here are some examples:  (i)~${\cal C}_2$ corresponds to a nearest-neighbor $XY$-coupling; (ii)~${\cal C}_4$ is associated with ring exchange terms such as in Eq.~(\protect\ref{ring}), {\em etc.}}   \label{Fig1}
\end{figure}
%%%%%%%%%%%%%%%%%%%%%%

We now can construct a general  spin model in two dimensions for which the particular transform (\ref{theferm}) directly applies. Let us reiterate that by ``applies,'' we mean  that the resulting fermion Hamiltonian does not
contain the $\tau$-operators or any phase factors arising from Jordan-Wigner strings. Otherwise, the transform (\ref{theferm}) as well as (\ref{JW}) always apply in the sense that they produce the fermion operators using the spin operators available. Let us write down the following rather general Hamiltonian associated with a square lattice
\begin{eqnarray}
\label{GenH}
\hat{\cal H} =  &&\!\!\!\! h_z\left[ \{ \hat{\sigma}_n^z \} \right] + \sum\limits_{n,l} \left[ \hat{\sigma}^\alpha(n,l) J_{\alpha \beta}(n,l) \hat{\sigma}^\beta(n+1,l)  \right]\\
&&\!\!\!\! +  \sum\limits_{{\cal C}  \subset \mbox{\small Even loops}} \sum\limits_{\{ a({\bf r}) =\pm \}}  J \left[\{ a({\bf r}) \}, {\cal C}\right] \prod_{{\bf r} \in  {\cal C}} \hat{\sigma}^{a({\bf r})} ({\bf r}) +{\rm h.~c.}
\nonumber
\end{eqnarray}
and decipher the meaning of each term in Eq.~(\ref{GenH}): The first term is an arbitrary function of $\hat{\sigma}^z$-operators on each site. The second term describes a collection of decoupled spin chains with arbitrary nearest-neighbor interactions within each chain. The last term includes all ``allowed'' interactions that involve two or more different chains (apart from those interactions, which may already be present in $h_z$). The indices $a({\bf r}) = \pm$ there label either lowering or raising operators, which may in principle appear in any combination, if no global spin conservation is imposed (no number conservation/gauge invariance for the fermions). Finally, ${\cal C}$ in Eq.~(\ref{GenH}) denotes a closed path, or a one-cycle, on a lattice:
${\cal C}_L =\left\{{\bf r}_1 \to {\bf r}_2  \to {\bf r}_3 \to \ldots \to {\bf r}_L \right\}$, where, ${\bf r}_i$, represents a lattice site and $\left|{\bf r}_{i+1} - {\bf r}_i\right| = \left|{\bf r}_{L} - {\bf r}_1\right| =0 \mbox{ or } 1$. The label ``Even loops''  in the sum correspond to the cycles that satisfy the following two constraints: (i)~Each even cycle has an even number of sites present from each leg and (ii)~For any site ${\bf r} = (n,l) \in {\cal C}$ lying in the path, there exists a ${\bf r}' = (n',l) \in {\cal C}$ from the same leg, which is either the same site ($n'=n$) or its nearest neighbor ($n=n'\pm1$).
 We note that including the first two terms in Eqs.~(\ref{GenH}) along with the general third term is in fact redundant, because the $\hat{\sigma}^z({\bf r})$-terms can be associated with ${\cal C}_z = \left\{ {\bf r} \to {\bf r} \right\}$ cycles  (see, Fig.~1) and the nearest-neighbor $XY$-interactions with  ${\cal C}_2 = \left\{ {\bf r} \to {\bf r} \pm {\bf e}_x \right\}$.   The purpose of all these complicated  constraints is simple: We want to avoid both $\tau$-operators in the resulting Hamiltonian [recall that $( \hat{\tau}^{x,y}_l )^{2n} = 1$, hence the choice of even cycles] and any phase factors that could arise from the  Jordan-Wigner string   (hence, we eliminate the non-nearest-neighbor interactions within legs~\cite{remark}). A general Hamiltonian (\ref{GenH}) satisfies all these conditions and can be fermionized via the proposed transform (\ref{theferm}). Fig.~1 illustrates the imposed constraints more clearly: The green loops there are examples of ``allowed'' paths, while the red loops provide examples of interaction terms, which are not allowed. Let us note that there may be a simple physical explanation why a Hamiltonian with terms of type $\overline{C}_8$ probably can not be fermionized via any variation of our method: The loop encircles a site that does not belong to it [$(n=8,l=6)$ in Fig.~1] and this site may or may not contain a fermion, which  leads to an ambiguity in the phase factors. Hence, the choice of statistics matters in this case, while for Hamiltonian (\ref{GenH}) it may become largely a question of convenience ({\em c.f.}, Ref.~[\onlinecite{VFL}]).

We can  generalize this scheme further  by considering more complicated long-range intra-chain interactions and various lattice structures including three-dimensional and higher-dimensional lattices (which  too can be conveniently fermionized using this method as long as the interaction terms respect the constraints, {\em i.e.}, belong to the allowed one- and higher-dimensional cycles). However instead of generalizing Hamiltonian~(\ref{GenH}), let us on the contrary focus on the following specific realization of it, which involves very interesting ring-exchange terms (associated with loops of type ${\cal C}_4$ in Fig.~1):
\begin{eqnarray}
\label{SLH}
\hat{\cal H}_{\rm SL} \!\! =  \!\!\!  \sum\limits_{{\bf r} \in \mathbb{Z}^2} \left\{
 \left[J_{||} \hat{\sigma}^+({\bf r}) \hat{\sigma}^-({\bf r} + \! {\bf e}_x\! ) +{\rm h.c.}\right] - B \hat{\sigma}^z({\bf r}) \right\} +\! \hat{\cal H}_z \! +\!   \hat{\cal H}_4,
\end{eqnarray}
where $\hat{\cal H}_z =  J_{z} \sum\limits_{{\bf r}}   \left[ \hat{\sigma}^z({\bf r}) \hat{\sigma}^z({\bf r}+{\bf e}_y) +{\rm h.c.}\right]$, {\em c.f.} Eq.~(\ref{XY2leg}), and
\begin{equation}
\label{ring}
\hat{\cal H}_4 = K_4 \sum\limits_{\bf r} \hat{\sigma}^+ ({\bf r}) \hat{\sigma}^- ({\bf r}+{\bf e}_x)
\hat{\sigma}^+ ({\bf r}+{\bf e}_x+{\bf e}_y)  \hat{\sigma}^- ({\bf r} + {\bf e}_y)   +{\rm h.~c.}
\end{equation}
is a ring-exchange term. Our method immediately gives the fermionized version of (\ref{SLH}) in the form of an array of Luttinger liquids coupled via density-density and current-current interactions. It is known that such a model hosts a variety of unusual phases, including so-called sliding Luttinger liquids~\cite{SLL1,SLL2,SLL3}. This immediately implies that the underlying spin model too remains a spin liquid in the corresponding parameter range.  Note that Eq.~(\ref{SLH}) is closely related to the model proposed earlier by Paramekanti, Balents, and Fisher~[\onlinecite{SL}] as a realization of a spin/Bose-liquid. Indeed,  the rotor Hamiltonian of Ref.~[\onlinecite{SL}],  ${\cal H} = U \sum\limits_{\bf r} \left( \hat{n}_{\bf r}  - \overline{n} \right)^2 - K  \sum\limits_{\bf r} \cos\left( \Delta_{xy} \hat{\phi}_{\bf r} \right)$, reduces in the hardcore-boson-limit to a spin-$(1/2)$ model of type (\ref{SLH}) with inter-chain hopping terms absent ({\em i.e.}, $J_{||} = 0$). Since the latter model can be fermionized as shown, the appearance of a liquid phase predicted in Ref.~[\onlinecite{SL}]  is quite natural. Let us also note that model (\ref{SLH}) is related to a Bose-metal phase introduced by Motrunich and Fisher~\cite{BM1}. The  proposed spin/Bose-metal Hamiltonian is of the following type $\hat{\cal H}_{\rm BM} = J \sum\limits_{\langle {\bf r}, {\bf r}' \rangle} \hat{\bm \sigma} ({\bf r}) \cdot \hat{\bm \sigma} ({\bf r}') + \hat{\cal H}_4$, where $\langle {\bf r}, {\bf r}' \rangle$ corresponds to nearest-neighbors on a square lattice or a spin-ladder~\cite{BM1,BM2}.  It is conceivable that the Bose metal may represent an isotropic version of the anisotropic spin liquid, which appeared in the array of $XY$-spin chains coupled by ring exchanges (\ref{SLH}). However, the presence of two-dimensional $XY$-couplings  in the generic Bose-metal Hamiltonian does not allow us to use the fermionization transform (\ref{theferm}).

Let us note that Jordan-Wigner-type fermionization with or without external operators, when applied to two- and higher-dimensional isotropic $XY$-models would produce a gauge theory for the fermions as opposed to a simple interacting theory, such as in Eq.~(\ref{2legferm}).  However, there may exist an alternative way to fermionize the $XY$-model in terms of BCS  fermions  with spin instead of spinless Jordan-Wigner fermions. It is known that the partition function of the Richardson model~\cite{Rich,VGR} of interacting fermions  with the Hamiltonian, $\hat{\cal H}_{\rm R} = \sum\limits_{{\bf r}, s =\uparrow,\downarrow} B_{\bf r} \hat{c}^\dagger_{{\bf r},s}  \hat{c}_{{\bf r},s} +  \sum\limits_{{\bf r},{\bf r}'} J({\bf r},{\bf r}') \hat{c}^\dagger_{{\bf r},\uparrow}  \hat{c}^\dagger_{{\bf r},\downarrow}  \hat{c}_{{\bf r}',\downarrow}  \hat{c}_{{\bf r}',\uparrow}$ can be mapped onto a combination partition functions, $Z_{\rm R} = \sum\limits \left( Z_{\rm spin} \times  Z_{\rm p-spin} \right)$, of real spins on singly-occupied sites and Anderson pseudo-spins~\cite{PA} on paired/empty sites, with $Z_{\rm p-spin}$ associated with an $XY$-model for Anderson pseudospins, $\hat{\cal H}_{\rm XY} =  \sum\limits_{{\bf r},{\bf r}'} J({\bf r},{\bf r}')\left[ \hat{\sigma}^+_{\bf r} \hat{\sigma}^-_{{\bf r}'} + {\rm h.c.}\right]$. Hence, by adding to Richardson Hamiltonian  non-linear terms that eliminate single occupancy, we can suppress the real-spin-sector and thereby fermionize the remaining $XY$-model associated with the pseudospins.

%%%%%%%%%%%%%%%%%%%%%%%%%%%%%
%\begin{figure}[t]
%\begin{center}
%\includegraphics[width=0.30\textwidth]{3leg_ladder.eps}
%\end{center}
%\caption{(Color online):~} An example of a three-leg spin ladder, which can be fermionized using a transform that involves just one ``external'' spin operator (see text for details) in contrast to  Eq.~(\ref{theferm}). \vspace*{-0.15in}  \label{Fig2}
%\end{figure}
%%%%%%%%%%%%%%%%%%%%%%%

In conclusion, let us mention that the general proposed scheme of using ``external'' operators to build a fermionized version of a spin model does not necessarily need to involve the particular transform (\ref{theferm}), which should be viewed as merely an example.
%The choice of the auxiliary operators is to be dictated entirely by convenience and after all these operators are needed only to prove the existence of a fermion counterpart to a given model. Once such a mapping is achieved, one generally can deal  with the fermion degrees of freedom only.
Furthermore, there may exist multiple different parameterizations or numberings of both the ``external'' operators and ``internal'' Jordan-Wigner-like strings that give rise to the same Hamiltonian and associated partition function (modulo an overall constant). To provide a final example in this context, let us consider a three-leg spin ladder, forming a triangular lattice. To fermionize this model, we in addition to transform (\ref{theferm}), can use, {\em e.g.}, the following fermion operators, $\hat{f}^\pm_l (n) = \hat{\sigma}^{\pm}(n,1) \prod_{m < n} \left[ \hat{\sigma}^{z}(m,1) \right] \hat{\tau}^l$, where $l=1,2,3 \equiv x,y,z$ is the leg index. Hence, we can fermionize this three-leg ladder using just one auxiliary spin, $\hat{\bm \tau}$. However, whether or not this particular parameterization is useful depends on a Hamiltonian of interest. If only nearest-leg ring exchanges are included, this transform leads to a proper fermion Hamiltonian, which does not involve the external spin. However, the inclusion of center-of-mass conserving ring-exchanges that involve all three legs (such as present in the models discussed in Refs.~[\onlinecite{BP},~\onlinecite{Burkov}]) would lead to the appearance of $\hat{\tau}^y$-``gauge''-factors in the fermion Hamiltonian.

This research was supported by NSF CAREER award DMR-0847224. The author acknowledges discussions with Anton Burkov and Tigran Sedrakyan.

\bibliography{myJW}

\begin{thebibliography}{22}
\expandafter\ifx\csname natexlab\endcsname\relax\def\natexlab#1{#1}\fi
\expandafter\ifx\csname bibnamefont\endcsname\relax
  \def\bibnamefont#1{#1}\fi
\expandafter\ifx\csname bibfnamefont\endcsname\relax
  \def\bibfnamefont#1{#1}\fi
\expandafter\ifx\csname citenamefont\endcsname\relax
  \def\citenamefont#1{#1}\fi
\expandafter\ifx\csname url\endcsname\relax
  \def\url#1{\texttt{#1}}\fi
\expandafter\ifx\csname urlprefix\endcsname\relax\def\urlprefix{URL }\fi
\providecommand{\bibinfo}[2]{#2}
\providecommand{\eprint}[2][]{\url{#2}}

\bibitem[{\citenamefont{Auerbach}()}]{AA_book}
\bibinfo{author}{\bibfnamefont{A.}~\bibnamefont{Auerbach}},
  \bibinfo{howpublished}{{\em ``Interacting Electrons and Quantum Magnetism,''}
  Springer-Verlag, New York, 1994}.

\bibitem[{\citenamefont{Giamarchi}()}]{Giamarchi_book}
\bibinfo{author}{\bibfnamefont{T.}~\bibnamefont{Giamarchi}},
  \bibinfo{howpublished}{{\em ``Quantum Physics in One Dimension,''} Oxford
  University Press, 2006}.

\bibitem[{\citenamefont{Jordan and Wigner}()}]{JW}
\bibinfo{author}{\bibfnamefont{P.}~\bibnamefont{Jordan}} \bibnamefont{and}
  \bibinfo{author}{\bibfnamefont{E.~P.} \bibnamefont{Wigner}},
  \bibinfo{howpublished}{Z. Phys. {\bf 47}, 631 (1928)}.

\bibitem[{\citenamefont{Fradkin}()}]{Fradkin_CS}
\bibinfo{author}{\bibfnamefont{E.}~\bibnamefont{Fradkin}},
  \bibinfo{howpublished}{Phys. Rev. Lett. {\bf 63}, 322 (1989)}.

\bibitem[{\citenamefont{Dai and bin Su}()}]{JW_ladders1}
\bibinfo{author}{\bibfnamefont{X.}~\bibnamefont{Dai}} \bibnamefont{and}
  \bibinfo{author}{\bibfnamefont{Z.}~\bibnamefont{bin Su}},
  \bibinfo{howpublished}{Phys. Rev. B {\bf 57}, 964 (1997)}.

\bibitem[{\citenamefont{Nunner and Kopp}()}]{JW_ladders2}
\bibinfo{author}{\bibfnamefont{T.~S.} \bibnamefont{Nunner}} \bibnamefont{and}
  \bibinfo{author}{\bibfnamefont{T.}~\bibnamefont{Kopp}},
  \bibinfo{howpublished}{Phys. Rev. B {\bf 69}, 104419 (2004)}.

\bibitem[{\citenamefont{Cheng et~al.}()\citenamefont{Cheng, Sun, Galitski, and
  {Das~Sarma}}}]{CSVGSDS}
\bibinfo{author}{\bibfnamefont{M.}~\bibnamefont{Cheng}},
  \bibinfo{author}{\bibfnamefont{K.}~\bibnamefont{Sun}},
  \bibinfo{author}{\bibfnamefont{V.}~\bibnamefont{Galitski}}, \bibnamefont{and}
  \bibinfo{author}{\bibfnamefont{S.}~\bibnamefont{{Das~Sarma}}},
  \bibinfo{howpublished}{Phys. Rev. B {\bf 81}, 024504 (2010)}.

\bibitem[{\citenamefont{Kitaev}()}]{Kitaev}
\bibinfo{author}{\bibfnamefont{A.~Y.} \bibnamefont{Kitaev}},
  \bibinfo{howpublished}{Usp. Fiz. Nauk (Suppl.) {\bf 171} (10), 131 (2001)}.

\bibitem[{\citenamefont{Jackiw and Rebbi}()}]{JR}
\bibinfo{author}{\bibfnamefont{R.}~\bibnamefont{Jackiw}} \bibnamefont{and}
  \bibinfo{author}{\bibfnamefont{C.}~\bibnamefont{Rebbi}},
  \bibinfo{howpublished}{Phys. Rev. D {\bf 13}, 3398 (1976)}.

\bibitem[{rem()}]{remark}
\bibinfo{howpublished}{Interaction terms that involve any long-range
  intra-chain couplings, but only short-range inter-chain couplings would still
  comply with the imposed requirements. However, such ``admissable'' long-range
  interactions are not included in model (\protect\ref{GenH}).}

\bibitem[{\citenamefont{Galitski et~al.}()\citenamefont{Galitski, Refael,
  Fisher, and Senthil}}]{VFL}
\bibinfo{author}{\bibfnamefont{V.~M.} \bibnamefont{Galitski}},
  \bibinfo{author}{\bibfnamefont{G.}~\bibnamefont{Refael}},
  \bibinfo{author}{\bibfnamefont{M.~P.~A.} \bibnamefont{Fisher}},
  \bibnamefont{and} \bibinfo{author}{\bibfnamefont{T.}~\bibnamefont{Senthil}},
  \bibinfo{howpublished}{Phys. Rev. Lett. {\bf 95}, 077002 (2005)}.

\bibitem[{\citenamefont{Emery et~al.}()\citenamefont{Emery, Fradkin, Kivelson,
  and Lubensky}}]{SLL1}
\bibinfo{author}{\bibfnamefont{V.~J.} \bibnamefont{Emery}},
  \bibinfo{author}{\bibfnamefont{E.}~\bibnamefont{Fradkin}},
  \bibinfo{author}{\bibfnamefont{S.~A.} \bibnamefont{Kivelson}},
  \bibnamefont{and} \bibinfo{author}{\bibfnamefont{T.~C.}
  \bibnamefont{Lubensky}}, \bibinfo{howpublished}{Phys. Rev. Lett. {\bf 85},
  2160 (2000)}.

\bibitem[{\citenamefont{Vishwanath and Carpentier}()}]{SLL2}
\bibinfo{author}{\bibfnamefont{A.}~\bibnamefont{Vishwanath}} \bibnamefont{and}
  \bibinfo{author}{\bibfnamefont{D.}~\bibnamefont{Carpentier}},
  \bibinfo{howpublished}{Phys. Rev. Lett. {\bf 86}, 676 (2001)}.

\bibitem[{\citenamefont{Mukhopadhyay et~al.}()\citenamefont{Mukhopadhyay, Kane,
  and Lubensky}}]{SLL3}
\bibinfo{author}{\bibfnamefont{R.}~\bibnamefont{Mukhopadhyay}},
  \bibinfo{author}{\bibfnamefont{C.~L.} \bibnamefont{Kane}}, \bibnamefont{and}
  \bibinfo{author}{\bibfnamefont{T.~C.} \bibnamefont{Lubensky}},
  \bibinfo{howpublished}{Phys. Rev. B {\bf 64}, 045120 (2001)}.

\bibitem[{\citenamefont{Paramekanti et~al.}()\citenamefont{Paramekanti,
  Balents, and Fisher}}]{SL}
\bibinfo{author}{\bibfnamefont{A.}~\bibnamefont{Paramekanti}},
  \bibinfo{author}{\bibfnamefont{L.}~\bibnamefont{Balents}}, \bibnamefont{and}
  \bibinfo{author}{\bibfnamefont{M.~P.~A.} \bibnamefont{Fisher}},
  \bibinfo{howpublished}{Phys. Rev. B {\bf 66}, 054526 (2002)}.

\bibitem[{\citenamefont{Motrunich and Fisher}()}]{BM1}
\bibinfo{author}{\bibfnamefont{O.~I.} \bibnamefont{Motrunich}}
  \bibnamefont{and} \bibinfo{author}{\bibfnamefont{M.~P.~A.}
  \bibnamefont{Fisher}}, \bibinfo{howpublished}{Phys. Rev. B {\bf 75}, 235116
  (2007)}.

\bibitem[{\citenamefont{Sheng et~al.}()\citenamefont{Sheng, Motrunich, and
  Fisher}}]{BM2}
\bibinfo{author}{\bibfnamefont{D.~N.} \bibnamefont{Sheng}},
  \bibinfo{author}{\bibfnamefont{O.~I.} \bibnamefont{Motrunich}},
  \bibnamefont{and} \bibinfo{author}{\bibfnamefont{M.~P.~A.}
  \bibnamefont{Fisher}}, \bibinfo{howpublished}{Phys. Rev. B {\bf 79}, 205112
  (2009)}.

\bibitem[{\citenamefont{Richardson}()}]{Rich}
\bibinfo{author}{\bibfnamefont{R.~W.} \bibnamefont{Richardson}},
  \bibinfo{howpublished}{Phys. Lett. {\bf 3}, 277 (1963); {\em ibid.} {\bf 5},
  82 (1963); R. W. Richardson and N. Sherman, Nucl. Phys. {\bf 52}, 221
  (1964)}.

\bibitem[{\citenamefont{Galitski}()}]{VGR}
\bibinfo{author}{\bibfnamefont{V.}~\bibnamefont{Galitski}},
  \bibinfo{howpublished}{arXiv:1003.2237v1 [cond-mat.supr-con] (2010)}.

\bibitem[{\citenamefont{Anderson}()}]{PA}
\bibinfo{author}{\bibfnamefont{P.~W.} \bibnamefont{Anderson}},
  \bibinfo{howpublished}{Phys. Rev. {\bf 112}, 1900 (1958)}.

\bibitem[{\citenamefont{Balents and Paramekanti}()}]{BP}
\bibinfo{author}{\bibfnamefont{L.}~\bibnamefont{Balents}} \bibnamefont{and}
  \bibinfo{author}{\bibfnamefont{A.}~\bibnamefont{Paramekanti}},
  \bibinfo{howpublished}{Phys. Rev. B {\bf 67}, 134427 (2003)}.

\bibitem[{\citenamefont{Burkov}()}]{Burkov}
\bibinfo{author}{\bibfnamefont{A.~A.} \bibnamefont{Burkov}},
  \bibinfo{howpublished}{Phys. Rev. B {\bf 81}, 125111 (2010)}.

\end{thebibliography}

\end{document}